\documentclass[aps,pra,twocolumn,superscriptaddress,10pt,twoside]{revtex4-2} 

\usepackage{amsmath,amssymb}
\usepackage{mathrsfs}
\usepackage{epstopdf}
\usepackage{graphicx}
\usepackage{bm,color,bbm}
\usepackage{natbib}
\usepackage[hidelinks]{hyperref}
\usepackage{caption}
\usepackage{subcaption}
\usepackage{fancyhdr}
\newcommand{\nn}{\nonumber}

\begin{document}
\fancyhead{} 
\fancyhead[LE,RO]{\ifnum\value{page}<2\relax\else\thepage\fi}

\title{Verifying Energy-Time Entanglement with Irregularly Sampled Correlations}

\author{James Schneeloch}
\email{james.schneeloch.1@afrl.af.mil}
\affiliation{Air Force Research Laboratory, Information Directorate, Rome, New York, 13441, USA}

\author{Christopher C. Tison}
\email{christopher.tison.2@afrl.af.mil}
\affiliation{Air Force Research Laboratory, Information Directorate, Rome, New York, 13441, USA}

\author{Richard J. Birrittella}
\affiliation{Booz Allen Hamilton, 8283 Greensboro Drive, McLean, VA 22102, USA}

\author{Ian Brinkley}
\affiliation{Center for Quantum Science and Engineering, University of California Las Angeles, Las Angeles, California, 90095}
\affiliation{UCLA Physics and Astronomy, PAB 4-437, 475 Portola Plaza, Los Angeles  California, 90095}
\author{Michael L. Fanto}
\affiliation{Air Force Research Laboratory, Information Directorate, Rome, New York, 13441, USA}
\affiliation{RIT Integrated Photonics group, Rochester Institute of Technology, Rochester, New York, 14623, USA}

\author{Paul M. Alsing}
\affiliation{Air Force Research Laboratory, Information Directorate, Rome, New York, 13441, USA}

\date{\today}

\begin{abstract}
Verifying entanglement with experimental measurements requires that we take the limitations of experimental techniques into account, while still proving that the data obtained could not have been generated from a classical source. In the energy-time degree of freedom, this challenge is exacerbated because realistic high-resolution frequency measurements are obtained as a function of light passing through arbitrary filters positioned at uneven intervals. In this work, we show how the data gathered from these kinds of measurements can be used to fully certify the degree of energy and timing correlations needed to certify energy-time entanglement without having to make special assumptions about the state or the measurement device. We accomplish this by showing how to construct a continuous-variable probability density from the data that can closely estimate, but never over-estimate the correlations (and entanglement) actually present in the system, and note that these methods are applicable to all continuous-variable degrees of freedom (e.g., spatial, field quadratures, etc). We illustrate the feasibility of these methods using frequency and timing correlations obtainable from photon pairs in Spontaneous Parametric Down-Conversion.
\end{abstract}


\maketitle
\thispagestyle{fancy}

\newpage

\section{Introduction}\label{SectionIntroduction}
As a means to quantum networking, energy-time entanglement between photons is a valuable resource because it may be transmitted over long distances through single-mode optical fiber with minimal distortion and loss. The amount of energy-time entanglement a pair of photons can share is virtually unlimited, but the direct characterization of it through measuring energy and timing correlations has remained a persistent challenge. While correlations in energy and time can be used to both witness and quantify entanglement when they exceed a classical threshold (as has been done successfully for position and momentum \cite{schneeloch2019quantifying}), the stringent energy and timing resolutions required have made this difficult to achieve directly.

To date, only one experiment \cite{Mei2020} has succeeded in directly demonstrating energy-time entanglement in this fashion, and that was because the photon pair source used was extremely narrowband using four-wave mixing in rubidium vapor. Using the quantification methods in \cite{SchneelochPra2018}, this four-wave mixing experiment effectively verifies an entanglement of formation of at least $2.54$ ebits per photon pair \footnote{With the reported value in \cite{Mei2020} of $\sigma(t_{A}-t_{B})\sigma(\omega_{A}+\omega_{B})=0.063\pm 0.0044$, and where \cite{SchneelochPra2018} shows that $-1$ minus the base-2 logarithm of this uncertainty product is a lower bound to the entanglement of formation in ebits, we obtain from their product a minimum value of $2.546\pm0.101$ ebits.}. Other demonstrations of energy-time entanglement have relied on extra assumptions about the state, such as assuming that the bandwidth of the frequency sum $\sigma(\omega_{A}+\omega_{B})$ must  equal that of the pump light consumed to generate the photon pairs (such as was done for three-photon energy-time entanglement in \cite{shalm2013three}), which can be measured separately to high precision. Alternatively, by assuming factors about the physics of the photon pair generation process (e.g., that the quantum and classical related nonlinear optical processes have related efficiencies), one can also determine the energy correlations to a high precision using stimulated emission tomography \cite{Liscidini_StimEmitTomog_PRL2013}. To overcome the timing resolution limits of conventional photon detectors, there has also been work in optical gating of the signal/idler photons with ultrashort pulses \cite{Macclean2018} where the enhanced timing resolution comes from an additional frequency conversion process  of the down-converted light that only occurs when the photons overlap with corresponding ultrashort pulses. 

In many cases, the correlations used to characterize entanglement are estimated by fitting functions to the data, and using the parameters of the fitting functions to define the degrees of correlation. Although this approach becomes more effective as the fit improves, it still risks overestimating correlations or mistakenly identifying entanglement in borderline cases. In this work, we show how to use such data to witness and certify entanglement without this manner of loophole.

The general strategy to determine energy-time entanglement directly through energy and timing correlations is to prove that these correlations exceed a threshold that all separable (i.e., unentangled) states obey. For energy and time (and other Fourier-conjugate pairs of observables), the most encompassing criterion thus far was developed by Walborn \emph{et.~al} in 2009 \cite{WalbornSepCrit2009}.

For Fourier-conjugate continuous-variable observables of two particles $A$ and $B$, Walborn \emph{et.~al} developed an entropy-based criterion that all separable states of $AB$ obey. Adapted for Fourier conjugates energy $E=\hbar\omega$ and (arrival) time $t$ the inequality is as follows:
\begin{equation}\label{WalbornEnergyTimeineq}
h(t_{A}-t_{B}) + h(\omega_{A}+\omega_{B})\geq\log(2\pi e).
\end{equation}
Here, $h(x)$ is the continuous Shannon entropy \cite{Cover2006} of probability density $\rho(x)$ obtained from the measurement outcomes of continuous observable $\hat{x}$, and all logarithms in this paper are assumed to be base-two since we measure entropy in bits. If we consider a state of minimum mixedness (i.e., a pure state), this inequality will witness entanglement for even the very smallest nonzero amounts of correlation \footnote{Indeed, one can lower-bound the amount of entanglement directly as the difference (when positive) between the measured correlations, and the  mixedness of the state described by its quantum entropy (as shown in equation 8 in \cite{schneeloch2023negativity}). For pure states (with zero quantum entropy) any nonzero degree of measured correlations places a corresponding nonzero minimum to the amount of entanglement.}. 
This makes this tool exceptionally powerful at identifying energy-time entanglement in the presence of both energy and time correlations.

For our analysis, we will use a looser entanglement witness, also developed by Walborn \emph{et.~al} \cite{WalbornEPRSteer2011} utilizing conditional quantum entropies as a measure of conditional uncertainty and correlation:
\begin{equation}\label{WalbornEnergyTimeEPR}
h(t_{A}|t_{B}) + h(\omega_{A}|\omega_{B})\geq\log(\pi e).
\end{equation}
where for example, the conditional entropy $h(t_{A}|t_{B})$ is equal to the difference between joint and marginal entropies: $h(t_{A},t_{B})-h(t_{B})$. This inequality \eqref{WalbornEnergyTimeEPR} is looser than \eqref{WalbornEnergyTimeineq} both because the bound on the right-hand side of \eqref{WalbornEnergyTimeEPR} is lower by one bit (i.e., $\log(2)$) than the right-hand side of \eqref{WalbornEnergyTimeineq}, and because the conditional entropies in \eqref{WalbornEnergyTimeEPR} are each lower bounds to the corresponding entropies of sums or differences in \eqref{WalbornEnergyTimeineq}.

Starting with Equation \eqref{WalbornEnergyTimeEPR} as our principal entanglement witness, we may witness entanglement (and also EPR-steering) whenever the conditional entropies $h(t_{A}|t_{B})$ and $h(\omega_{A}|\omega_{B})$ add to a value smaller than $\log(\pi e)$. Determining this experimentally is a challenge because one cannot obtain a completely faithful representation of the probability density $\rho(\omega_{A},\omega_{B})$ with a finite number of measurements carried out at finite resolution. To resolve this challenge, it was shown in \cite{PhysRevLett.110.130407} and subsequently in \cite{schneeloch2015demonstrating, schneeloch2019quantifying} how measurements of continuous observables carried out at finite resolution can still be used to place tight upper bounds to these conditional entropies. This is distinct from the case of constructing new uncertainty relations for discrete measurements of continuous observables (as reviewed in \cite{e20060454}).

Where entropy is a measure of mixedness \cite{schneeloch2023negativity}, and subsequent mixing (i.e., majorization) operations cannot decrease the entropy or increase correlation, this paper is focused on how to obtain a conservative estimate of the correlations present from experimental data. This process follows multiple steps of increasing complexity: 
\begin{itemize}
    \item First, we review how the simple coarse-graining of a continuous-variable (CV) probability distribution $\rho(\omega_{A},\omega_{B})$ into pixels of finite width $\Delta\omega$ and transmission spectra $\Pi(\omega-\omega_{n})$, equally spaced with center frequencies $\omega_{n}=\omega_{0}+n\Delta\omega$ (what we call simple-top-hat coarse graining) creates a distribution $\bar{\rho}_{\Pi}(\omega_{A},\omega_{B})$ whose continuous conditional entropy cannot be lower than that of the underlying CV probability distribution $\rho(\omega_{A},\omega_{B})$.
    \item Next, we consider that high-resolution frequency measurements are carried out by examining the fraction of biphotons that pass through a pair of tuneable-center frequency filters whose transmission spectra $f(\omega-\omega_{n})$ differ from the top-hat distributions $\Pi(\omega-\omega_{n})$ (generally looking more like Lorentzian functions). We prove that when the spacing $\Delta\omega$ between two adjacent filter positions $(\omega_{n},\omega_{n+1})$ is small enough that a top-hat distribution of the same width would majorize these filter functions, then the resulting probability distribution $\bar{\rho}_{f}(\omega_{A},\omega_{B})$ has a conditional entropy no less than that obtained from $\bar{\rho}_{\Pi}(\omega_{A},\omega_{B})$ via the simple top-hat coarse graining, and subsequently, also than that from the underlying CV probability distribution $\rho(\omega_{A},\omega_{B})$.
    \item Third, we consider the more general case where these filter functions not only differ from the top-hat function, but also from one another, where they are only approximately regularly spaced, and only approximately equal in shape to one another due to experimental imperfections in the measurement apparatus. Modulo a shift of frequency center, we show that the extent to which these filter functions $f_{n}(\omega-\omega_{n})$ are equal to one another, one can determine and apply a correction factor to the conditional entropy to ensure that the subsequent statistics never underestimate the entropy, nor over-estimate entanglement.
\end{itemize} 

The rest of the paper is laid out as follows.  For background, we first describe the concept of majorization for both discrete and continuous probability distributions and discuss its consequences in the estimation of entropy and correlations. We then detail the preceding measurement strategies, showing how to obtain conservative estimates of the entropies and correlations of the underlying quantum observables. To conclude this work, we illustrate the effectiveness of our techniques by showing how typical data obtained from measuring energy and timing correlations between photon pairs generated in type-I spontaneous parametric down-conversion (SPDC) in periodically poled potassium titanyl phosphate (PPKTP) could be used to verify entanglement.

\section{Majorization, entropy, and entanglement}
Majorization of discrete probability distributions is a well-used tool to order different probability distributions by mixedness (when possible). Given two probability distributions $\vec{p}$ and $\vec{q}$ (whose components sum to unity), we say that $\vec{p}$ \emph{majorizes} $\vec{q}$ (denoted $\vec{p}\succ \vec{q}$), when the sum of the $n$ largest elements of $\vec{p}$ is greater than or equal to the sum of the $n$ largest elements of $\vec{q}$ for all $n$ from $1$ to the dimension of $\vec{p}$ and $\vec{q}$ \footnote{Note: Because their components sum to unity, majorization and weak majorization between two probability vectors are equivalent.}.

When $\vec{p}\succ \vec{q}$, one can obtain $\vec{q}$ from $\vec{p}$ through a series of permutations and mixing operations where pairs of outcomes with unequal probabilities are brought closer to their arithmetic mean. All measures of mixedness $f(\vec{p})$ \footnote{We define for a probability distribution $\vec{p}$ that a measure of mixedness $f(\vec{p})$ is a Schur-concave function whose minimum value is obtained when one element of $\vec{p}$ equals unity with all others being zero. This minimum value is most often set equal to zero.}(such as entropy, among others) will agree that if $\vec{p}\succ \vec{q}$, then $f(\vec{p})\leq f(\vec{q})$.

Majorization may be used to witness entanglement by violating the majorization criterion \cite{Nielsen2001}, which witnesses entanglement whenever a joint state is less mixed than its marginal subsystems (by any measure of mixedness)\cite{schneeloch2023negativity}. However, this requires knowledge of the eigenspectrum of the joint and marginal density matrices, which becomes prohibitively difficult to acquire at high dimension. Instead, we may also use majorization to witness entanglement by violating entanglement criteria that place limits on the relative mixedness of the classical probability distributions of measured observables.

To connect majorization with the conservative estimation of correlations (and subsequently entanglement), we use the finding in the supplemental material of \cite{schneeloch2019quantifying}, which shows that for any two measurement probability distributions $\vec{p}$ and $\vec{q}$ (joint, marginal, or otherwise), the relative entropy between them (denoted as $\mathscr{D}(\vec{p}||\vec{q})$) cannot increase following a majorization operation (where the same operation is applied to both $\vec{p}$ and $\vec{q}$). Because the joint entropy, marginal entropy, conditional entropy, and mutual information can each be expressed in terms of relative entropies between pairs of probability distributions \cite{Cover2006}, one can show that majorization operations cannot decrease entropies of any kind, nor increase the mutual information. This is useful because if we can use experimental data to determine probability distributions for time and frequency that are definitely majorized by the true underlying distributions, one obtains upper limits to the entropies in the left hand side of the inequality \eqref{WalbornEnergyTimeEPR}, which when totalling to less than $\log(\pi e)$, will witness energy-time entanglement.


\section{Coarse graining at different levels of complexity}

In general, our task is to determine probability distributions consistent with measurement data that are provably majorized by the unknown underlying probability distributions to prove that the underlying correlations must exceed an entanglement-witnessing threshold. There are multiple strategies to do this, each of which is progressively more challenging.

\subsection{Case 0: Simple top-hat coarse graining $\Pi(\omega-\omega_{n})$} 
Top-hat coarse-graining involves partitioning a continuous space of outcomes (e.g., frequency) into discrete intervals, and using the discrete probabilities corresponding to the fraction of events occurring within each interval to create a probability density $\bar{\rho}_{\Pi}(\omega_{A},\omega_{B})$ majorized by the underlying statistics $\rho(\omega_{A},\omega_{B})$. This works because the total probability occurring within an interval is equal to integrating the mean probability density over that same interval. Where replacing the underlying probability density with its mean values over each interval is a form of mixing/majorization, simple top-hat coarse-graining yields probability distributions that are majorized by the underlying observable statistics. As a result, the entropies and correlations we obtain from $\bar{\rho}_{\Pi}(\omega_{A},\omega_{B})$ can be used to certify entanglement without over-estimating it. For example, we have:
\begin{equation}
H_{\Pi}(\Omega_{A}|\Omega_{B}) + \log(\Delta\omega_{A}) = h_{\Pi}(\omega_{A}|\omega_{B})\geq h(\omega_{A}|\omega_{B})
\end{equation}
where $H_{\Pi}(\Omega_{A}|\Omega_{B})$ is the conditional entropy of the discrete measurement probabilities of photon $A$ having frequency within the bin $\Omega_{A}$, given that photon $B$ is known to be within frequency bin $\Omega_{B}$ and the measurement probabilities are obtained by integrating over simple top-hat functions denoted by $\Pi$. In addition, $\Delta\omega_{A}$ is the size of the frequency bins of $\omega_{A}$, and $h_{\Pi}(\omega_{A}|\omega_{B})$ is the continuous conditional entropy of the discretized probability densities $\bar{\rho}_{\Pi}(\omega_{A},\omega_{B})$ and $\bar{\rho}_{\Pi}(\omega_{B})$. For a sufficiently high resolution (i.e., sufficiently small values of $\Delta\omega_{A}$ and $\Delta\omega_{B}$), the value of $H(\Omega_{A}|\Omega_{B}) +\log(\Delta\omega_{A})$ will converge (from above) to the continuous-conditional entropy $h(\omega_{A}|\omega_{B})$.

For frequency measurements, this can be achieved by measuring the total power reflected off of a diffraction grating in discrete intervals, though this becomes technically challenging at the required frequency resolutions and ranges. For timing measurements, this is more straightforward, though one is still limited by the timing resolution of one's photon counters. In practice, detector jitter and tagger jitter may be dominant sources of uncertainty in the timing measurements.

\subsection{Case 1: filter-shifted coarse graining $f(\omega-\omega_{n})$}
Where simple coarse graining measurements may be impractical, it is still possible to acquire frequency information by examining how much light passes through a given frequency filter as a function of its center frequency.

\begin{figure}[t]
\centerline{\includegraphics[width=0.85\columnwidth]{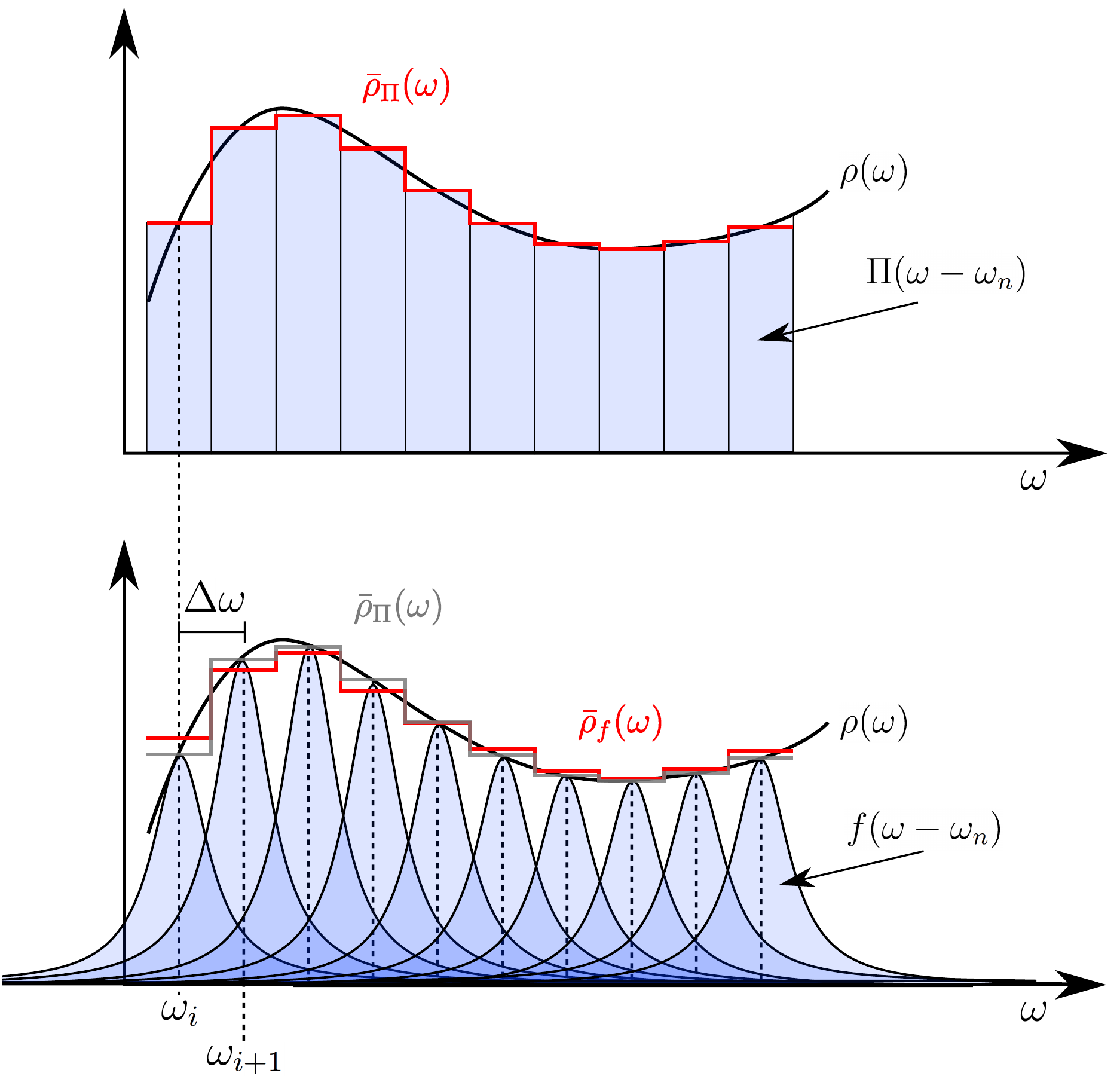}}
\caption{Diagram showing difference between top-hat coarse graining, and sampling with a regularly spaced Lorentzian filter. The area under the shaded curves or rectangles roughly corresponds to the discrete probability associated to that frequency spacing. When the width of the Lorentzian filters $f(\omega-\omega_{n})$ is wide enough that it is majorized by the corresponding top-hat filter $\Pi(\omega-\omega_{n})$, then the corresponding discretized probability density $\bar{\rho}_{f}(\omega)$ is guaranteed to majorize $\rho(\omega)$.}\label{CaseDiagram}
\end{figure}

The case of simple-coarse graining can be contextualized as recording the light passing through a top-hat filter $\Pi(\omega-\omega_{n})$, positioned at frequency intervals equal to its width $\Delta\omega$ (so that we have central frequencies $\omega_{n}=\omega_{0}+n\Delta\omega$). The top-hat coarse-grained probability density $\bar{\rho}_{\Pi}(\omega)$ is then equal to $\rho(\omega)$ but where within each interval of size $\Delta\omega$, $\rho(\omega)$ is replaced by its mean value over the same interval. Because the light passing though a top-hat filter is (up to a constant factor) equal to the amount of light incident on a discrete bin in frequency probability space, this method is already known to provide the statistics needed to verify entanglement. See Fig.~\ref{CaseDiagram} for diagram.

The complexity arises when we consider filters with shapes other than a top-hat function. In particular, narrow-linewidth frequency filters (such as from a Fabry-Perot etalon) used for higher-resolution frequency measurements have an approximately Lorentzian rather than top-hat transmission spectrum. In what follows, we prove that for a fixed (albeit arbitrary) filter profile $f(\omega-\omega_{n})$, sampled at equal intervals, we can also obtain the statistics needed to verify entanglement.

If we consider equally spaced normalized frequency filters given by $f(\omega-\omega_{n})$ (such that it integrates to unity), then whenever the peak of $f(\omega-\omega_{n})$ is lower than $1/\Delta\omega$ (i.e., the peak of the normalized top-hat filter $\Pi(\omega-\omega_{n})$),  the resulting coarse-grained probability distribution $\bar{\rho}_{f}(\omega_{A},\omega_{B})$ will be majorized by what can be obtained with simple top-hat course graining $\bar{\rho}_{\Pi}(\omega_{A},\omega_{B})$. 

To illustrate this, we provide the following quirk of majorization. One probability distribution majorizes another if and only if the other distribution can be expressed as the result of applying a doubly-stochastic operator on the first. For distributions $f(\omega-\omega_{n})$ and $\Pi(\omega-\omega_{n})$ and any central frequency $\omega_{n}$, we can assert
\begin{subequations}
\begin{align}
f(\omega-\omega_{n})&\prec \Pi(\omega-\omega_{n})\\
\text{if and only if } f(\omega-\omega_{n})&=\!\!\int\!\! d\omega' p(\omega,\omega')\Pi(\omega'-\omega_{n})
\end{align}
\end{subequations}
where $p(\omega,\omega')$ is a doubly-stochastic function (such that its range is non-negative, and integrating over either $\omega$ or $\omega'$ gives unity). Moreover, because $p(\omega,\omega')$ is independent of $\omega_{n}$, it may be taken to be a convolution kernel such that $f(\omega-\omega_{n})$ is the convolution of $\Pi(\omega-\omega_{n})$ and an unknown but fixed distribution $p(\omega)$.

With the majorization condition defined, the discrete measurement probabilities $P(\Pi_{n})$ and $P(f_{n})$ of continuous probability density $\rho(\omega)$ are given by:
\begin{subequations}\label{majRel}
\begin{align}
P(\Pi_{n})&=\mathcal{N}_{\Pi}\int d\omega \rho(\omega) \Pi(\omega-\omega_{n})\\
P(f_{n})&=\mathcal{N}_{f}\int d\omega \rho(\omega) f(\omega-\omega_{n})
\end{align}
\end{subequations}
and the discretized probability densities $\bar{\rho}_{\pi}(\omega)$ and $\bar{\rho}_{f}(\omega)$ are given by:
\begin{subequations}\label{majRel2}
\begin{align}
\bar{\rho}_{\Pi}(\omega)&=\sum_{n}P(\Pi_{n})\Pi(\omega-\omega_{n})\\
\bar{\rho}_{f}(\omega)&=\sum_{n}P(f_{n})\Pi(\omega-\omega_{n})
\end{align}
\end{subequations}
where the normalization constants $\mathcal{N}_{\Pi}$ and $\mathcal{N}_{f}$ are defined such that the discrete probabilities summed over $n$ add to unity. 

Another property of majorization we shall make use of is its transitivity (i.e., if $p\prec q$ and $q\prec r$, then $p\prec r$). In particular, we shall show that when $f(\omega-\omega_{n})$ is majorized by $\Pi(\omega-\omega_{n})$, that $\bar{\rho}_{f}(\omega)$ is majorized  by $\rho(\omega)$ because it is majorized by a different probability density $\bar{\rho}(\omega)$ which is itself majorized by $\rho(\omega)$.

To begin, we examine the discrete probabilities $P(f_{n})$ from which $\bar{\rho}_{f}(\omega)$ is constructed on the assumption that $f(\omega-\omega_{n})\prec \Pi(\omega-\omega_{n})$.
\begin{subequations}
\begin{align}
P(f_{n})&=\mathcal{N}_{f}\!\!\int\!\! d\omega \rho(\omega) f(\omega-\omega_{n})\\
&=\mathcal{N}_{f}\!\!\int\!\! d\omega \rho(\omega) \left(\!\!\int\!\! d\omega' p(\omega,\omega')\Pi(\omega'-\omega_{n})\right)\\
&=\mathcal{N}_{f}\!\!\iint\!\! d\omega' d\omega \;\rho(\omega)p(\omega,\omega')\Pi(\omega'-\omega_{n})\\
&=\mathcal{N}_{f}\!\!\int\!\! d\omega' \left(\!\!\int\!\! d\omega \rho(\omega)p(\omega,\omega')\right)\Pi(\omega'-\omega_{n})\\
&=\mathcal{N}_{f}\!\!\int\!\! d\omega' \bar{\rho}(\omega')\Pi(\omega'-\omega_{n})
\end{align}
\end{subequations}
The first simplification comes from the majorization relationship allowing us to substitute $f(\omega-\omega_{n})$ for the doubly-stochastic integral over $\Pi(\omega-\omega_{n})$. By regrouping terms, we can see that these probabilities $P(f_{n})$ can be expressed as a simple-top-hat coarse graining over a new probability density $\bar{\rho}(\omega)$, also expressed as a doubly-stochastic integral over $\rho(\omega)$. From this we understand first, that $\bar{\rho}(\omega)\prec \rho(\omega)$, and that $\bar{\rho}_{f}(\omega)\prec \bar{\rho}(\omega)$. Therefore, we can conclude that $\bar{\rho}_{f}(\omega)\prec \rho(\omega)$, and the statistics from these frequency measurements can be used to place conservative bounds on the entropies used in the entanglement witness \eqref{WalbornEnergyTimeEPR} so as not to falsely witness it. In particular, we may state that if $f(\omega-\omega_{n})\prec \Pi(\omega-\omega_{n})$, then
\begin{equation}
H_{f}(\Omega_{A}|\Omega_{B})+\log(\Delta\omega_{A})\geq h(\omega_{A}|\omega_{B}).
\end{equation}
Where $H_{f}(\Omega_{A}|\Omega_{B})$ is the discrete conditional entropy obtained from the measurement probabilities gathered from light passing through filters $f(\omega-\omega_{n})$ instead of $\Pi(\omega-\omega_{n})$.

As a quick example, we point out that given a frequency spacing of $\Delta\omega$: a scanning Lorentzian filter would need a FWHM $a$ greater than $(2/\pi)\Delta\omega$ (about $0.637\Delta\omega$); and a scanning Gaussian filter would need a standard deviation $\sigma_{\omega}$ greater than $\Delta\omega/\sqrt{2\pi}$ (or about $0.399\Delta\omega$). Filters between these two profiles, such as a Lorentzian with Gaussian jitter sampled over time (i.e. a Voigt profile) will have a similar characteristic width, and a correspondingly similar sampling spacing requirement to the Lorentzian and Gaussian filter cases. In general, we may expect that a filter whose width is larger than the spacing $\Delta\omega$ between sampled frequencies will be one that is majorized by the corresponding simple top-hat filter.

\subsection{Case 2: Sampling with changing filters\\at changing intervals}\label{hardestcase}
Thus far, we have a viable strategy for conservatively characterizing frequency correlations between photon pairs on the assumption that we assemble a coarse-grained probability distribution by shifting a pair of filters to different pairs of frequencies (such that the spacing between adjacent filter positions is generally a good deal smaller than its linewidth to satisfy the majorization relation). To make the preceding methods more fully general, we have to consider that the filter $f(\omega-\omega_{n})$ will generally change shape when set to different central frequencies (however slightly). In what follows, we document how bounding the difference between the fluctuating frequency filter $f_{n}(\omega-\omega_{n})$ and an empirical ideal $f(\omega-\omega_{n})$ can be used to establish a conservative estimate of correlations even when the filtering process is not a static one.

If we relax the assumption that the filter functions $f_{n}(\omega-\omega_{n})$ have an identical profile (modulo a shift), then the doubly-stochastic operator will itself be a function of $n$:
\begin{subequations}
\begin{align}
f_{n}(\omega\!-\!\omega_{n})&=\!\!\int\!\! d\omega' p_{n}(\omega,\omega')\Pi(\omega'-\omega_{n})\\
P(f_{n})&=\mathcal{N}_{f}\!\!\!\!\int\!\! d\omega' \!\!\left(\!\!\int\!\! d\omega \rho(\omega)p_{n}(\omega,\omega')\!\!\right)\!\!\Pi(\omega'\!\!-\!\omega_{n})\\
&=\mathcal{N}_{f}\!\!\int\!\! d\omega' \bar{\rho}_{n}(\omega')\Pi(\omega'\!\!-\omega_{n})
\end{align}
\end{subequations}
where here, the distributions $\bar{\rho}_{n}(\omega)$ each are majorized by $\rho(\omega)$, but it remains to be proven whether the statistics of $P(f_{n})$ in this situation can be used to conservatively estimate entanglement.

\begin{figure}[t]
\centerline{\includegraphics[width=0.85\columnwidth]{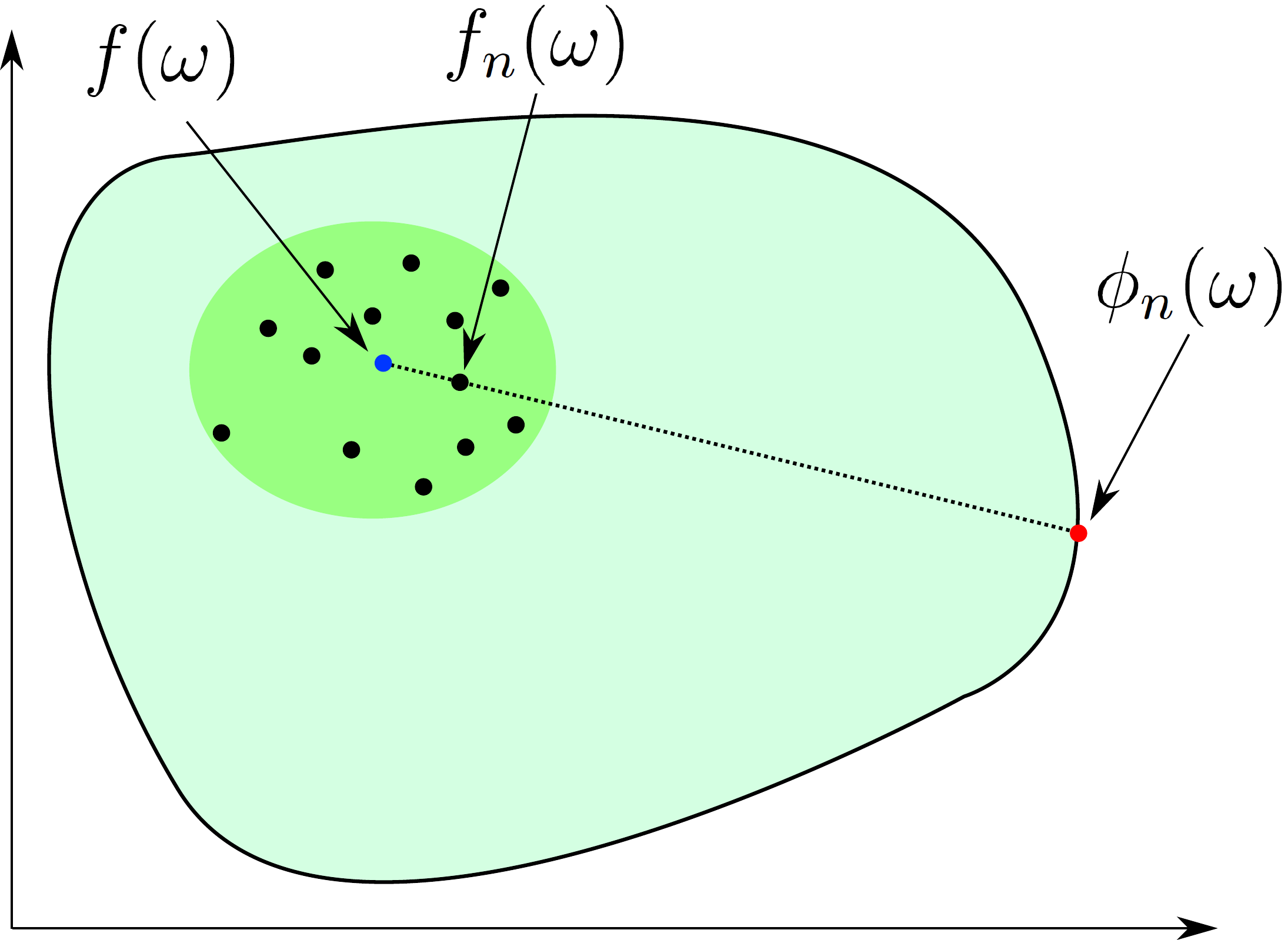}}
\caption{Diagram showing convex set normalized filter functions and how the individual filter functions $f_{n}(\omega)$ relate to the target filter function $f(\omega)$ and extremal filter function $\phi_{n}(\omega)$. In particular, $f(\omega)$ is taken as the arithmetic mean of $f_{n}(\omega)$ over all values of $n$.}\label{Dscochastic}
\end{figure}

Let us consider the situation where all of these filters are ``nearly" equivalent, such that the normalized filter functions $f_{n}(\omega-\omega_{n})$ are all nearly equal (for the same center frequency) to a single filter function $f(\omega-\omega_{n})$. Accounting for this same central frequency, we would say $f_{n}(\omega)\approx f(\omega)$. Because the set of normalized filter functions is convex, every filter function $f_{n}(\omega)$ can be expressed as a weighted mixture of a fixed filter function $f(\omega)$, and an extremal filter function $\phi_{n}(\omega)$:
\begin{equation}\label{weightedaverage}
f_{n}(\omega)= w_{n} f(\omega) + (1-w_{n})\phi_{n}(\omega)
\end{equation}
To the extent that all of the filter functions $f_{n}(\omega)$ are nearly equal to one another, we can select a target filter function $f(\omega)$ such the weights $w_{n}$ are on average nearly unity. See Fig.~\ref{Dscochastic} for diagram. In particular, by taking $f(\omega)$ to be the arithmetic mean of all $f_{n}(\omega)$, this target function is one whose mean square distance to all functions $f_{n}(\omega)$ is minimized.

Within the convex set of normalized filter functions, we define the extremal filter functions $\phi_{n}(\omega)$ relative to $f(\omega)$ and $f_{n}(\omega)$ such that a line starting at $f(\omega)$ and passing through $f_{n}(\omega)$ will intersect the boundary of the set at $\phi_{n}(\omega)$. See Fig.~\ref{Dscochastic} for diagram.

Using this weighted decomposition of $f_{n}(\omega)$, and the fact that the simple top-hat coarse graining is a linear operation, one can decompose the discrete probability distribution $P(f_{n})$ similarly using the same set of weights:
\begin{align}
P(f_{n}) &= w_{n} P(f_{n}|p) + (1-w_{n})P(f_{n}|\phi_{n})
\end{align}
Here, $P(f_{n}|p)$ can be associated with the probability of passing through filter $f_{n}$ given it is drawn from the target function $f(\omega)$, and $P(f_{n}|\phi_{n})$ is similarly defined, relative to the extremal filter function $\phi_{n}(\omega)$.

Extending this analysis to two dimensions (so as to measure joint statistics for biphotons passing through pairs of filters), we have:
\begin{equation}
    P(f_{m},f_{n}) = w_{mn} P(f_{m},f_{n}|p) + (1-w_{mn})P(f_{m},f_{n}|\phi_{mn})
\end{equation}
In what follows, we shall show how a minimum weight $w_{0}$, defining the relative closeness of different filter functions $f_{n}(\omega)$, can be used as a correction factor between the conditional entropies $H(f_{A}|f_{B})$ (obtained through measurement) and $H_{f}(\Omega_{A}|\Omega_{B})$ (assuming a constant filter function) so that its resulting value remains a conservative estimation.

Let $P(G=1|m,n)$ be the probability that the statistics are drawn from the target distribution when employing filter types $(m,n)$ (also given by weight $w_{mn}$), and let $P(G=0|m,n)$ be the corresponding probability that this does not happen. Then the joint frequency measurement distribution $P(f_{A},f_{B}|m,n)$ (also given as $P(f_{m},f_{n})$) when using filter types $m$ and $n$ can be broken up according to these weighted probabilities:
\begin{align}
P(f_{A},f_{B}|m,n)&=P(G=1|m,n)P(f_{A},f_{B}|G=1,m,n) \nonumber\\
&\!\!\!\!\!\!\!\!\!\!\!\!\!\!\!+P(G=0|m,n)P(f_{A},f_{B}|G=0,m,n)
 \end{align}
Then, assuming that the target filter function is majorized by the top-hat distribution defined by the frequency bin sizes, we have:
\begin{equation}
H(f_{A}|f_{B},G=1)\geq H_{f}(\Omega_{A}|\Omega_{B}).
\end{equation}

From this, we can make the following relation regarding the conditional entropy $H(f_{A}|f_{B})$:
\begin{align}
H(f_{A}|f_{B})&\geq \sum_{m,n} P(m,n) H(f_{A}|f_{B},G,m,n)\nn\\
&\!\!\!\!\!\!\!\!\!\!\!\!\!\!\!\!\!\!\!\!= \sum_{m,n} P(m,n)\bigg[ P(G=0|m,n)H(f_{A}|f_{B},G=0,m,n) \nn\\
&+ P(G=1|m,n)H(f_{A}|f_{B},G=1,m,n)\bigg]\nn\\
&\!\!\!\!\!\!\!\!\!\!\!\!\!\!\!\!\!\!\!\!\geq \sum_{m,n}P(m,n)P(G=1|m,n)H(f_{A}|f_{B},G=1,m,n)\nn\\
&\!\!\!\!\!\!\!\!\!\!\!\!\!\!\!\!\!\!\!\!=\bigg(\sum_{m,n} P(m,n)P(G=1|m,n)\bigg)H_{f}(\Omega_{A}|\Omega_{B})\nn\\
&\!\!\!\!\!\!\!\!\!\!\!\!\!\!\!\!\!\!\!\!\geq \bigg(\!\!\min_{m,n}(P(G=1|m,n)) \!\!\bigg) H_{f}(\Omega_{A}|\Omega_{B})\nn\\
&\!\!\!\!\!\!\!\!\!\!\!\!\!\!\!\!\!\!\!\!=w_{0} \;H_{f}(\Omega_{A}|\Omega_{B})
\end{align}
The simplification between the second and third expressions comes from the non-negativity of the entropy. The simplification from the third to the fourth expression comes from the statement that the filter function specified by $G=1$ is independent of values $m$ and $n$. Between the fourth and fifth expression, we have used the fact that the mean value of a function must be greater than or equal to its minimum value.

It is in this way, that when $P(G=1)$ is close to unity, that we may conservatively put a useful upper bound on $h(\omega_{A}|\omega_{B})$:
\begin{align}
\frac{H(f_{A}|f_{B})}{w_{0}} +\log(\Delta\omega_{A})&\geq H_{f}(\Omega_{A}|\Omega_{B})+\log(\Delta\omega_{A})\nn\\
&\geq h(\omega_{B}|\omega_{A}).
\end{align}
Thus, by determining a minimum value of $w_{0}$, one may witness energy-time entanglement so long as the frequency spacing  is sufficiently narrow relative to the filter linewidth.

As the filter functions $f_{n}(\omega)$ can be measured independently, we can explicitly determine $w_{0}$ by considering their proximity to the mean filter function, or some other common centroid. Given how the filter functions $f_{n}(\omega)$ are expressible as the weighted average \eqref{weightedaverage} between a target filter function $f(\omega)$ and extremal filter function $\phi_{n}(\omega)$, the filter function $f_{n}(\omega)$ is always intermediate between $f(\omega)$ and $\phi_{n}(\omega)$ for all values of $\omega$ (See Fig.~\ref{Dscochastic} for diagram). These quantities (in addition to $w_{n}$) are all non-negative, which together gives us the inequality:
\begin{equation}
w_{n}\leq\frac{f_{n}(\omega)}{f(\omega)}
\end{equation}
for all values of $\omega$. Then, because $\phi_{n}(\omega)$ is on the boundary of the set of possible probability density functions, there must be at least one value $\omega_{0}$ for which $\phi_{n}(\omega_{0})=0$, but where $f(\omega_{0})\neq 0$ and $f_{n}(\omega_{0})\neq 0$. Our inequality is thus tight, and we have a formula for $w_{n}$ in terms of determined or measured quantities:
\begin{equation}
w_{n}=\min_{\omega}\bigg(\frac{f_{n}(\omega)}{f(\omega)}\bigg).
\end{equation}
Once this weight is determined for each filter function, the final value for $w_{0}$ will just be the minimum among these weights.

Note: For these weights to be usable, the filter function cannot decay too rapidly (i.e., exponentially or faster) because for finite changes in the parameters of rapidly decaying filter functions, there will be values of $\omega$ for which the ratio/weight $w_{n}$ will be arbitrarily close to zero. Fortunately, the most common frequency filters have Lorentzian spectra (decaying more slowly), which may be well-utilized in this style of measurement.

As one example, we can consider a target Lorentzian filter function $f(\omega)$ of FWHM $a$, and the empirical filter function $f_{n}(\omega)$ of $FWHM$ $a(1+\epsilon_{n})$ (for $\epsilon_{n}\geq 0$). In this case, $w_{n}$ would be $1/(1+\epsilon_{n})$, and for small $\epsilon_{n}$, would differ from unity by approximately $\epsilon_{n}$ as well.

For a Lorentzian filter function with a constant $FWHM$ of $a$, but a small varying central position differing from $x_{n}$ by amount $\delta_{n} a$, we have:
\begin{equation}
w_{n}=1+\frac{\delta_{n}^{2}}{2}-\delta_{n}\sqrt{1+\frac{\delta_{n}^{2}}{4}},
\end{equation}
which for small deviations, approaches:
\begin{equation}
w_{n}\approx 1 -\delta_{n},
\end{equation}
Thus, for Lorentzian functions, this measurement strategy can be expected to work well so long as the deviations of the functions from the target are a relatively small fraction of the FWHM.

\section{Suitability of majorization-based sampling using typical experimental parameters}

The general experiment we consider requires a source of energy-time entangled photon pairs, but is not subject to any one physical process of generating them. For ease of illustration, we consider the process of Spontaneous Parametric Down-Conversion (SPDC) to be what generates these entangled biphotons, but it is not the only kind of process that can do so. For our experiment, we consider a single-mode pump laser shining through a $\chi^{(2)}$-nonlinear-optical medium (typically a noncentrosymmetric crystal). Inside this medium, provided the conditions are met for simultaneous energy and momentum conservation (i.e., phase matching) events will occur in which single pump photons are annihilated to create signal-idler biphotons (known as SPDC). These photon pairs can exhibit high levels of energy-time entanglement because their point of origin is nearly simultaneous (so that $\sigma(t_{A}-t_{B})$ will be small) implying large timing correlations; and their energies must sum to the energy of the annihilated pump photon (so that $\sigma(\omega_{A}+\omega_{B})$ is small for narrow pump bandwidths) implying large energy correlations. As mentioned earlier in this work, when these correlations together exceed a critical threshold \eqref{WalbornEnergyTimeEPR}, we may directly witness the biphoton's energy-time entanglement.

\subsection{Timing correlation acquisition}
In this subsection, we will show how the timing correlations of photon pairs are measured experimentally using coincidence counting. In short, coincidence counting allows us to obtain an empirical probability distribution of the arrival time difference $t_{A}-t_{B}$ by compiling a histogram of these timing differences between two distinct photon detectors.

To record the number of photon pairs being generated with single-photon detectors, we split the biphotons so that one photon goes into one arm, and the other photon goes into another arm toward respective single-photon detectors. When the first detector registers a photon, a photon correlator used to measure coincidence counts starts a timer, which is stopped only when the second detector registers an event. As photons hit these detectors, a histogram of these time intervals builds up such that pairs of photons generated simultaneously will have a characteristic peak in this coincidence histogram associated to the difference in time each photon takes to travel from its source to its detector. As these timing measurements are digital, the coincidence peak covers a range of time bins, and one may use this discrete data of the empirical probability distribution of $(t_{A}-t_{B})$ to conservatively bound the entropies used in witnessing energy-time entanglement via simple top-hat course graining.


After finding the coincidence window, and subtracting off the relatively small background, we may normalize the resulting histogram quantized into bins of size $\Delta t$ into an empirical discrete probability distribution, and use it to conservatively bound the continuous entropy $h(t_{A}-t_{B})$ with the corresponding entropy of its discrete approximation:
\begin{equation}
h(t_{A}-t_{B}) \leq H(T_{A}-T_{B}) + \log(\Delta t) 
\end{equation}
Here, $\Delta t$ is set by the time bin size chosen to make the coincidence histogram (in this case, 1 ps).

Beyond the time-bin size, this method of measuring $t_{A}-t_{B}$ is limited in precision by the jitter time of each detector, which corresponds to the uncertainty between when a photon is absorbed, and when the detector generates an electrical pulse corresponding to a single photon count. With detector jitter times at around $20$ picoseconds, and assuming their jitters are gaussian-distributed and statistically independent, this means that if the photon pairs were completely simultaneous, the minimum width of the coincidence histogram would be $28.28$ps, or the jitter time multiplied by $\sqrt{2}$.

Independent of technical noise, there is still a fundamental nonzero time uncertainty (of $t_{A}-t_{B}$) of the photon pairs as determined by the physics of SPDC, and in particular, by the phase matching. For a type-0 or type-I degenerate SPDC source, we may estimate \cite{schneeloch2016introduction} this uncertainty as:
\begin{equation}
\sigma(t_{A}-t_{B}) \approx \sqrt{\frac{9 l_{z}\kappa_{1}}{10}}\end{equation}
which for a $20$mm crystal of PPKTP pumped at $775$nm,  (where $\kappa$, the group velocity dispersion constant is determined from Sellmeier equations in \cite{kato2002sellmeier} to be $292 fs^2/mm$) may be estimated to be of the order $72.5$fs, which assuming a Gaussian distribution corresponds to a FWHM of about $171.$fs. This is well-below the jitter time uncertainty of $2.828\times10^{4}$ fs, so these pairs may be regarded as nearly simultaneous.

While we would use the empirical probabilities to experimentally bound the continuous entropy $h(t_{A}-t_{B})$ by its discrete approximation, we can also use the given parameters to estimate how much entanglement may be witnessed for these biphotons. As a Gaussian distribution has maximum entropy for a given mean and variance, we may upper-bound the entropy $h(t_{A}-t_{B})$ as:
\begin{align}
h(t_{A}-t_{B})&\leq \frac{1}{2}\log\left(2\pi e\sigma(t_{A}-t_{B})^{2}\right)\\
&\approx \frac{1}{2}\log\left(\frac{\pi e (FWHM(t_{A}-t_{B}))^2}{4\ln(2)}\right)
\end{align}

Using our example $FWHM(t_{A}-t_{B})$ of $424$ps, and the timing resolution was $1$ps, we find the continuous time entropy to be approximately $-30.342$ bits. This defines the threshold of energy correlations we must surpass to demonstrate energy-time entanglement.

\subsection{Energy correlation acquisition}
\subsubsection{The limit to beat}
Given our maximal value of $-30.324$ bits for $h(t_{A}-t_{B})$, and the uncertainty bound of $\log(\pi e)\approx 3.0942$ bits, this means that $h(\omega_{A}+\omega_{B})$ must be less than $33.42$ bits, corresponding to a  $\sigma(\omega_{A}+\omega_{B})$ less than $2\pi(442.$ MHz), or when centered at $1550$nm, a linewidth (FWHM) of $8.8$pm.

The joint frequency distribution of the biphotons from SPDC has an uncertainty $\sigma(\omega_{A}+\omega_{B})$ dominated by the pump bandwidth, which may be arbitrarily narrow. Consequently, our ability to demonstrate entanglement will be dominated by the resolutions we may achieve in different frequency measurement setups.

\subsubsection{Direct quantization into frequency bins via diffraction or dispersion}
The most straightforward way to perform these joint frequency measurements is to split the signal and idler photons into separate arms, use a spatially dispersive element such as diffraction grating or prism in concert with other optics including a digital micromirror device (DMD) array, to reflect portions of the joint frequency spectrum toward or away from photon detectors from which coincidence counts could be recorded. With this manner of acquisition and the previously acquired timing correlations, we would need frequency bins small enough to resolve the linewidth. Where the discrete approximation to the continuous entropy is an upper bound to it, $H(\Omega_{A}+\Omega_{B}) +\log(\Delta\omega) \geq h(\omega_{A}+\omega_{B})$, we need this discrete entropy approximation to total a value less than $33.42$ bits. Where the bin size places a lower limit on the FWHM, we would need frequency bins significantly smaller than $1$ GHz, or $8$pm (ideally by an order of magnitude) in order to begin witnessing entanglement. In what follows, we describe the technological limitations of different optical frequency measurement techniques.

Diffraction gratings are rated by their resolving power and diffraction efficiency. Where the diffraction efficiency describes the fraction of input light that exits the first diffracted mode, the resolving power is given from the Rayleigh criterion of being able to just barely distinguish two spectral lines in the far field. A typical diffraction grating adapted to the telecom band might have $600$ grooves per mm and be able to have a $10$ mm diameter beam incident upon it without clipping. Together, this corresponds to a resolving power of about one part in $6000$ for the beam of light. If our central frequency is $193.4$ THz (i.e., $1550$nm), then one might expect to be able to resolve the difference between two spectral lines $32$ GHz apart. Any finer spectral features would effectively be blurred to approximately this size. As the maximum frequency bin size needed to witness energy-time entanglement (with the same timing correlation data) is around an order of magnitude below $1$ GHz, any apparatus utilizing a diffraction grating will have to be impractically large to have a sufficient resolving power (e.g., a $320-3200$ mm wide diffraction grating uniformly illuminated by a single beam).

Prisms may be rated by their resolving power as well. \cite{Ramsay:40} \footnote{For an equilateral prism of side length $b$, the Rayleigh resolving power of a prism (surrounded by vacuum) may be given by the formula $\Delta\lambda/\lambda \approx -b n(n-n_{g})/\lambda$.}. For typical equilateral triangle dispersive prisms, their resolving power is competitive with what can be accomplished with a diffraction grating, though this may be greatly enhanced by using a material with very high dispersion (by as much as three orders of magnitude \cite{AtomicPrismPRA2012}), though this dispersion often exists only at a narrow range of frequencies such as between two atomic transitions in a triangular rubidium vapor cell. Unlike diffraction gratings, ordinary prisms may work over a broader range of wavelengths, but this comes at the expense of having to keep track of how the refractive index varies over a large range of wavelengths (which may be far from a linear function over such a range). 

Where both diffraction gratings and prisms rely on wavelength-dependent beam deflection to make spectral measurements possible, there is a fundamental tradeoff between the minimum resolvable difference between two wavelengths, and the maximum range of wavelengths over which data can be acquired, as these correspond to minimum and maximum observable angular deflections. This is also true for additional methods relying on angular dispersion, such as Virtually Imaged Phased Arrays (VIPAs) \cite{VIPAPatent, xiao2004dispersion}  which have much improved frequency resolution albeit over a much narrower range. These arrays can be concatenated with other optical dispersive elements to achieve high resolutions over a comparatively wide bandwidth \cite{xiao20042, Zhu_2020}, though it remains to be shown how this method of measurement can be used to conservatively estimate entanglement. Fortunately, there are alternative methods to separate different frequency bands of light simple enough that their analysis toward realizing energy-time entanglement is straightforward. In what follows, we consider the feasibility of acquiring the necessary frequency statistics by scanning tunable narrow-linewidth frequency filters.

\subsubsection{Requirements of scanning filters}

For our frequency measurement, the last situation we consider is scanning a pair of tuneable Fabry-Perot Etalons. Where the pump spectrum may be arbitrarily narrow, we would be limited to the width of the transmission peaks of these etalon filters, which may be described by a Lorentzian function. With a Lorentzian filter profile (so that $h(\omega_{1}+\omega_{2})=\log(2\pi FWHM(\omega_{1}+\omega_{2}))$), we find that the maximum value for the $FWHM(\omega_{1}+\omega_{2})$ is $2\pi\times 0.29$GHz. For scale, at $1550$nm, a fluctuation of $1.25GHz$ is a fluctuation of $0.01nm$, so these resolutions, while challenging to achieve, are obtainable with current technology.

The most dramatic advantage in using tunable frequency filters over other frequency sampling techniques (not counting higher resolution) is that one can perform a continuous scan of an arbitrary curve in the joint frequency space to trace out lines over which $\omega_{A}+\omega_{B}$ is a constant, forming a direct measurement of the probability distribution of $\omega_{A}+\omega_{B}$ whose entropy may be calculated and used to more sensitively witness energy-time entanglement by violating \eqref{WalbornEnergyTimeineq} instead of \eqref{WalbornEnergyTimeEPR}. When the correlations are sufficiently strong that we can violate the entanglement witness by a non-negligible margin, we may use the methods in section \ref{hardestcase} to compensate for the intrinsic variability in the filter function scanned at different frequencies, due to technical, electronic, and mechanical noise in the measurement process.

\section{Discussion: Relative necessity of different-case sampling approaches}
At this point, we have discussed three classes of methods of increasing complexity to discretely sample continuous observables to conservatively bound their continuous entropies and in so doing, conservatively witness continuous-variable entanglement. At sufficiently high sensitivity, all experimental measurements are subject to fluctuations, so it may seem like the strongest, most complex and resource-intensive approach is always necessary to demonstrate energy-time entanglement, but this need not be the case in practical situations.

The joint frequency probability distributions obtained from counting the number of biphotons passing through a pair of frequency filters as a function of those filters' centers is subject to shot-noise (Poisson-distributed) uncertainty in the number of detected biphotons. Where these coincidence counting rates fluctuate, so do the corresponding conservative estimates of probabilities, entropies, correlation, and entanglement. As is done in the estimation of other physical parameters, we need not consider every conceivable influence that contributes to the experimental data if that effect can be shown to be sufficiently small.

Although the most conservative acquisition strategy (in which we measure all of the filter functions at all center frequencies to gauge their relative uniformity and apply a correction factor to the measured entropy) is valid in all situations, it also becomes significantly resource-intensive at high measurement resolution over a broad range. However, if the effect to the measured entropies from the 
variation/jitter of these filter functions  cannot be resolved from the competing effect of shot noise, the statistics we obtain will essentially be unperturbed, and we may opt for the simpler approach.

\section{Conclusion}
Until recently, there has been no direct demonstration of energy-time entanglement through energy and time correlations except for extremely narrowband sources where techniques exist to scan the limited bandwidth directly at high resolution. Until now, the less standard techniques required for high-resolution frequency measurement made rigorous demonstrations of energy-time entanglement a puzzling endeavor. We have shown how these kinds of measurements may still be used to rigorously demonstrate entanglement, and explored the technological landscape required to achieve it experimentally. In demonstrating how both to directly and conservatively sample these correlations, we make it possible to utilize energy-time entanglement without making extra assumptions about the system or measurement device that might otherwise compromise their utility in quantum networking.

\begin{acknowledgments}
We gratefully acknowledge support from the Air Force Office of Scientific Research as well as insightful comments from Dr. Dylan Heberle and Dr. David Hucul.

The views expressed are those of the authors and do not reflect the official guidance or position of the United States Government, the Department of Defense or of the United States Air Force. The appearance of external hyperlinks does not constitute endorsement by the United States Department of Defense (DoD) of the linked websites, or of the information, products, or services contained therein. The DoD does not exercise any editorial, security, or other control over the information you may find at these locations.
\end{acknowledgments}

\bibliography{EPRbib16}

\begin{thebibliography}{27}%
\makeatletter
\providecommand \@ifxundefined [1]{%
 \@ifx{#1\undefined}
}%
\providecommand \@ifnum [1]{%
 \ifnum #1\expandafter \@firstoftwo
 \else \expandafter \@secondoftwo
 \fi
}%
\providecommand \@ifx [1]{%
 \ifx #1\expandafter \@firstoftwo
 \else \expandafter \@secondoftwo
 \fi
}%
\providecommand \natexlab [1]{#1}%
\providecommand \enquote  [1]{``#1''}%
\providecommand \bibnamefont  [1]{#1}%
\providecommand \bibfnamefont [1]{#1}%
\providecommand \citenamefont [1]{#1}%
\providecommand \href@noop [0]{\@secondoftwo}%
\providecommand \href [0]{\begingroup \@sanitize@url \@href}%
\providecommand \@href[1]{\@@startlink{#1}\@@href}%
\providecommand \@@href[1]{\endgroup#1\@@endlink}%
\providecommand \@sanitize@url [0]{\catcode `\\12\catcode `\$12\catcode `\&12\catcode `\#12\catcode `\^12\catcode `\_12\catcode `\%12\relax}%
\providecommand \@@startlink[1]{}%
\providecommand \@@endlink[0]{}%
\providecommand \url  [0]{\begingroup\@sanitize@url \@url }%
\providecommand \@url [1]{\endgroup\@href {#1}{\urlprefix }}%
\providecommand \urlprefix  [0]{URL }%
\providecommand \Eprint [0]{\href }%
\providecommand \doibase [0]{https://doi.org/}%
\providecommand \selectlanguage [0]{\@gobble}%
\providecommand \bibinfo  [0]{\@secondoftwo}%
\providecommand \bibfield  [0]{\@secondoftwo}%
\providecommand \translation [1]{[#1]}%
\providecommand \BibitemOpen [0]{}%
\providecommand \bibitemStop [0]{}%
\providecommand \bibitemNoStop [0]{.\EOS\space}%
\providecommand \EOS [0]{\spacefactor3000\relax}%
\providecommand \BibitemShut  [1]{\csname bibitem#1\endcsname}%
\let\auto@bib@innerbib\@empty
\bibitem [{\citenamefont {Schneeloch}\ \emph {et~al.}(2019)\citenamefont {Schneeloch}, \citenamefont {Tison}, \citenamefont {Fanto}, \citenamefont {Alsing},\ and\ \citenamefont {Howland}}]{schneeloch2019quantifying}%
  \BibitemOpen
  \bibfield  {author} {\bibinfo {author} {\bibfnamefont {J.}~\bibnamefont {Schneeloch}}, \bibinfo {author} {\bibfnamefont {C.~C.}\ \bibnamefont {Tison}}, \bibinfo {author} {\bibfnamefont {M.~L.}\ \bibnamefont {Fanto}}, \bibinfo {author} {\bibfnamefont {P.~M.}\ \bibnamefont {Alsing}},\ and\ \bibinfo {author} {\bibfnamefont {G.~A.}\ \bibnamefont {Howland}},\ }\bibfield  {title} {\bibinfo {title} {Quantifying entanglement in a 68-billion-dimensional quantum state space},\ }\href@noop {} {\bibfield  {journal} {\bibinfo  {journal} {Nature communications}\ }\textbf {\bibinfo {volume} {10}},\ \bibinfo {pages} {1} (\bibinfo {year} {2019})}\BibitemShut {NoStop}%
\bibitem [{\citenamefont {Mei}\ \emph {et~al.}(2020)\citenamefont {Mei}, \citenamefont {Zhou}, \citenamefont {Zhang}, \citenamefont {Li}, \citenamefont {Liao}, \citenamefont {Yan}, \citenamefont {Zhu},\ and\ \citenamefont {Du}}]{Mei2020}%
  \BibitemOpen
  \bibfield  {author} {\bibinfo {author} {\bibfnamefont {Y.}~\bibnamefont {Mei}}, \bibinfo {author} {\bibfnamefont {Y.}~\bibnamefont {Zhou}}, \bibinfo {author} {\bibfnamefont {S.}~\bibnamefont {Zhang}}, \bibinfo {author} {\bibfnamefont {J.}~\bibnamefont {Li}}, \bibinfo {author} {\bibfnamefont {K.}~\bibnamefont {Liao}}, \bibinfo {author} {\bibfnamefont {H.}~\bibnamefont {Yan}}, \bibinfo {author} {\bibfnamefont {S.-L.}\ \bibnamefont {Zhu}},\ and\ \bibinfo {author} {\bibfnamefont {S.}~\bibnamefont {Du}},\ }\bibfield  {title} {\bibinfo {title} {Einstein-podolsky-rosen energy-time entanglement of narrow-band biphotons},\ }\href {https://doi.org/10.1103/PhysRevLett.124.010509} {\bibfield  {journal} {\bibinfo  {journal} {Phys. Rev. Lett.}\ }\textbf {\bibinfo {volume} {124}},\ \bibinfo {pages} {010509} (\bibinfo {year} {2020})}\BibitemShut {NoStop}%
\bibitem [{\citenamefont {Schneeloch}\ and\ \citenamefont {Howland}(2018)}]{SchneelochPra2018}%
  \BibitemOpen
  \bibfield  {author} {\bibinfo {author} {\bibfnamefont {J.}~\bibnamefont {Schneeloch}}\ and\ \bibinfo {author} {\bibfnamefont {G.~A.}\ \bibnamefont {Howland}},\ }\bibfield  {title} {\bibinfo {title} {{Quantifying high-dimensional entanglement with Einstein-Podolsky-Rosen correlations}},\ }\href {https://doi.org/10.1103/PhysRevA.97.042338} {\bibfield  {journal} {\bibinfo  {journal} {Phys. Rev. A}\ }\textbf {\bibinfo {volume} {97}},\ \bibinfo {pages} {042338} (\bibinfo {year} {2018})}\BibitemShut {NoStop}%
\bibitem [{Note1()}]{Note1}%
  \BibitemOpen
  \bibinfo {note} {With the reported value in \cite {Mei2020} of $\sigma (t_{A}-t_{B})\sigma (\omega _{A}+\omega _{B})=0.063\pm 0.0044$, and where \cite {SchneelochPra2018} shows that $-1$ minus the base-2 logarithm of this uncertainty product is a lower bound to the entanglement of formation in ebits, we obtain from their product a minimum value of $2.546\pm 0.101$ ebits.}\BibitemShut {Stop}%
\bibitem [{\citenamefont {Shalm}\ \emph {et~al.}(2013)\citenamefont {Shalm}, \citenamefont {Hamel}, \citenamefont {Yan}, \citenamefont {Simon}, \citenamefont {Resch},\ and\ \citenamefont {Jennewein}}]{shalm2013three}%
  \BibitemOpen
  \bibfield  {author} {\bibinfo {author} {\bibfnamefont {L.~K.}\ \bibnamefont {Shalm}}, \bibinfo {author} {\bibfnamefont {D.~R.}\ \bibnamefont {Hamel}}, \bibinfo {author} {\bibfnamefont {Z.}~\bibnamefont {Yan}}, \bibinfo {author} {\bibfnamefont {C.}~\bibnamefont {Simon}}, \bibinfo {author} {\bibfnamefont {K.~J.}\ \bibnamefont {Resch}},\ and\ \bibinfo {author} {\bibfnamefont {T.}~\bibnamefont {Jennewein}},\ }\bibfield  {title} {\bibinfo {title} {Three-photon energy--time entanglement},\ }\href@noop {} {\bibfield  {journal} {\bibinfo  {journal} {Nature Physics}\ }\textbf {\bibinfo {volume} {9}},\ \bibinfo {pages} {19} (\bibinfo {year} {2013})}\BibitemShut {NoStop}%
\bibitem [{\citenamefont {Liscidini}\ and\ \citenamefont {Sipe}(2013)}]{Liscidini_StimEmitTomog_PRL2013}%
  \BibitemOpen
  \bibfield  {author} {\bibinfo {author} {\bibfnamefont {M.}~\bibnamefont {Liscidini}}\ and\ \bibinfo {author} {\bibfnamefont {J.~E.}\ \bibnamefont {Sipe}},\ }\bibfield  {title} {\bibinfo {title} {Stimulated emission tomography},\ }\href {https://doi.org/10.1103/PhysRevLett.111.193602} {\bibfield  {journal} {\bibinfo  {journal} {Phys. Rev. Lett.}\ }\textbf {\bibinfo {volume} {111}},\ \bibinfo {pages} {193602} (\bibinfo {year} {2013})}\BibitemShut {NoStop}%
\bibitem [{\citenamefont {MacLean}\ \emph {et~al.}(2018)\citenamefont {MacLean}, \citenamefont {Donohue},\ and\ \citenamefont {Resch}}]{Macclean2018}%
  \BibitemOpen
  \bibfield  {author} {\bibinfo {author} {\bibfnamefont {J.-P.~W.}\ \bibnamefont {MacLean}}, \bibinfo {author} {\bibfnamefont {J.~M.}\ \bibnamefont {Donohue}},\ and\ \bibinfo {author} {\bibfnamefont {K.~J.}\ \bibnamefont {Resch}},\ }\bibfield  {title} {\bibinfo {title} {Direct characterization of ultrafast energy-time entangled photon pairs},\ }\href {https://doi.org/10.1103/PhysRevLett.120.053601} {\bibfield  {journal} {\bibinfo  {journal} {Phys. Rev. Lett.}\ }\textbf {\bibinfo {volume} {120}},\ \bibinfo {pages} {053601} (\bibinfo {year} {2018})}\BibitemShut {NoStop}%
\bibitem [{\citenamefont {Walborn}\ \emph {et~al.}(2009)\citenamefont {Walborn}, \citenamefont {Taketani}, \citenamefont {Salles}, \citenamefont {Toscano},\ and\ \citenamefont {de~Matos~Filho}}]{WalbornSepCrit2009}%
  \BibitemOpen
  \bibfield  {author} {\bibinfo {author} {\bibfnamefont {S.~P.}\ \bibnamefont {Walborn}}, \bibinfo {author} {\bibfnamefont {B.~G.}\ \bibnamefont {Taketani}}, \bibinfo {author} {\bibfnamefont {A.}~\bibnamefont {Salles}}, \bibinfo {author} {\bibfnamefont {F.}~\bibnamefont {Toscano}},\ and\ \bibinfo {author} {\bibfnamefont {R.~L.}\ \bibnamefont {de~Matos~Filho}},\ }\bibfield  {title} {\bibinfo {title} {Entropic entanglement criteria for continuous variables},\ }\href {https://doi.org/10.1103/PhysRevLett.103.160505} {\bibfield  {journal} {\bibinfo  {journal} {Phys. Rev. Lett.}\ }\textbf {\bibinfo {volume} {103}},\ \bibinfo {pages} {160505} (\bibinfo {year} {2009})}\BibitemShut {NoStop}%
\bibitem [{\citenamefont {Cover}\ and\ \citenamefont {Thomas}(2006)}]{Cover2006}%
  \BibitemOpen
  \bibfield  {author} {\bibinfo {author} {\bibfnamefont {T.~M.}\ \bibnamefont {Cover}}\ and\ \bibinfo {author} {\bibfnamefont {J.~A.}\ \bibnamefont {Thomas}},\ }\href@noop {} {\emph {\bibinfo {title} {{Elements of Information Theory}}}},\ \bibinfo {edition} {2nd}\ ed.\ (\bibinfo  {publisher} {Wiley and Sons},\ \bibinfo {address} {New York},\ \bibinfo {year} {2006})\BibitemShut {NoStop}%
\bibitem [{Note2()}]{Note2}%
  \BibitemOpen
  \bibinfo {note} {Indeed, one can lower-bound the amount of entanglement directly as the difference (when positive) between the measured correlations, and the mixedness of the state described by its quantum entropy (as shown in equation 8 in \cite {schneeloch2023negativity}). For pure states (with zero quantum entropy) any nonzero degree of measured correlations places a corresponding nonzero minimum to the amount of entanglement.}\BibitemShut {Stop}%
\bibitem [{\citenamefont {Walborn}\ \emph {et~al.}(2011)\citenamefont {Walborn}, \citenamefont {Salles}, \citenamefont {Gomes}, \citenamefont {Toscano},\ and\ \citenamefont {Souto~Ribeiro}}]{WalbornEPRSteer2011}%
  \BibitemOpen
  \bibfield  {author} {\bibinfo {author} {\bibfnamefont {S.~P.}\ \bibnamefont {Walborn}}, \bibinfo {author} {\bibfnamefont {A.}~\bibnamefont {Salles}}, \bibinfo {author} {\bibfnamefont {R.~M.}\ \bibnamefont {Gomes}}, \bibinfo {author} {\bibfnamefont {F.}~\bibnamefont {Toscano}},\ and\ \bibinfo {author} {\bibfnamefont {P.~H.}\ \bibnamefont {Souto~Ribeiro}},\ }\bibfield  {title} {\bibinfo {title} {Revealing hidden einstein-podolsky-rosen nonlocality},\ }\href {https://doi.org/10.1103/PhysRevLett.106.130402} {\bibfield  {journal} {\bibinfo  {journal} {Phys. Rev. Lett.}\ }\textbf {\bibinfo {volume} {106}},\ \bibinfo {pages} {130402} (\bibinfo {year} {2011})}\BibitemShut {NoStop}%
\bibitem [{\citenamefont {Schneeloch}\ \emph {et~al.}(2013)\citenamefont {Schneeloch}, \citenamefont {Dixon}, \citenamefont {Howland}, \citenamefont {Broadbent},\ and\ \citenamefont {Howell}}]{PhysRevLett.110.130407}%
  \BibitemOpen
  \bibfield  {author} {\bibinfo {author} {\bibfnamefont {J.}~\bibnamefont {Schneeloch}}, \bibinfo {author} {\bibfnamefont {P.~B.}\ \bibnamefont {Dixon}}, \bibinfo {author} {\bibfnamefont {G.~A.}\ \bibnamefont {Howland}}, \bibinfo {author} {\bibfnamefont {C.~J.}\ \bibnamefont {Broadbent}},\ and\ \bibinfo {author} {\bibfnamefont {J.~C.}\ \bibnamefont {Howell}},\ }\bibfield  {title} {\bibinfo {title} {Violation of continuous-variable einstein-podolsky-rosen steering with discrete measurements},\ }\href {https://doi.org/10.1103/PhysRevLett.110.130407} {\bibfield  {journal} {\bibinfo  {journal} {Phys. Rev. Lett.}\ }\textbf {\bibinfo {volume} {110}},\ \bibinfo {pages} {130407} (\bibinfo {year} {2013})}\BibitemShut {NoStop}%
\bibitem [{\citenamefont {Schneeloch}\ \emph {et~al.}(2015)\citenamefont {Schneeloch}, \citenamefont {Knarr}, \citenamefont {Howland},\ and\ \citenamefont {Howell}}]{schneeloch2015demonstrating}%
  \BibitemOpen
  \bibfield  {author} {\bibinfo {author} {\bibfnamefont {J.}~\bibnamefont {Schneeloch}}, \bibinfo {author} {\bibfnamefont {S.~H.}\ \bibnamefont {Knarr}}, \bibinfo {author} {\bibfnamefont {G.~A.}\ \bibnamefont {Howland}},\ and\ \bibinfo {author} {\bibfnamefont {J.~C.}\ \bibnamefont {Howell}},\ }\bibfield  {title} {\bibinfo {title} {Demonstrating continuous variable einstein--podolsky--rosen steering in spite of finite experimental capabilities using fano steering bounds},\ }\href@noop {} {\bibfield  {journal} {\bibinfo  {journal} {JOSA B}\ }\textbf {\bibinfo {volume} {32}},\ \bibinfo {pages} {A8} (\bibinfo {year} {2015})}\BibitemShut {NoStop}%
\bibitem [{\citenamefont {Toscano}\ \emph {et~al.}(2018)\citenamefont {Toscano}, \citenamefont {Tasca}, \citenamefont {Rudnicki},\ and\ \citenamefont {Walborn}}]{e20060454}%
  \BibitemOpen
  \bibfield  {author} {\bibinfo {author} {\bibfnamefont {F.}~\bibnamefont {Toscano}}, \bibinfo {author} {\bibfnamefont {D.~S.}\ \bibnamefont {Tasca}}, \bibinfo {author} {\bibfnamefont {L.}~\bibnamefont {Rudnicki}},\ and\ \bibinfo {author} {\bibfnamefont {S.~P.}\ \bibnamefont {Walborn}},\ }\bibfield  {title} {\bibinfo {title} {Uncertainty relations for coarse–grained measurements: An overview},\ }\bibfield  {journal} {\bibinfo  {journal} {Entropy}\ }\textbf {\bibinfo {volume} {20}},\ \href {https://doi.org/10.3390/e20060454} {10.3390/e20060454} (\bibinfo {year} {2018})\BibitemShut {NoStop}%
\bibitem [{\citenamefont {Schneeloch}\ \emph {et~al.}(2023)\citenamefont {Schneeloch}, \citenamefont {Tison}, \citenamefont {Jacinto},\ and\ \citenamefont {Alsing}}]{schneeloch2023negativity}%
  \BibitemOpen
  \bibfield  {author} {\bibinfo {author} {\bibfnamefont {J.}~\bibnamefont {Schneeloch}}, \bibinfo {author} {\bibfnamefont {C.~C.}\ \bibnamefont {Tison}}, \bibinfo {author} {\bibfnamefont {H.~S.}\ \bibnamefont {Jacinto}},\ and\ \bibinfo {author} {\bibfnamefont {P.~M.}\ \bibnamefont {Alsing}},\ }\bibfield  {title} {\bibinfo {title} {Negativity vs. purity and entropy in witnessing entanglement},\ }\href@noop {} {\bibfield  {journal} {\bibinfo  {journal} {Scientific Reports}\ }\textbf {\bibinfo {volume} {13}},\ \bibinfo {pages} {4601} (\bibinfo {year} {2023})}\BibitemShut {NoStop}%
\bibitem [{Note3()}]{Note3}%
  \BibitemOpen
  \bibinfo {note} {Note: Because their components sum to unity, majorization and weak majorization between two probability vectors are equivalent.}\BibitemShut {Stop}%
\bibitem [{Note4()}]{Note4}%
  \BibitemOpen
  \bibinfo {note} {We define for a probability distribution $\protect \vec {p}$ that a measure of mixedness $f(\protect \vec {p})$ is a Schur-concave function whose minimum value is obtained when one element of $\protect \vec {p}$ equals unity with all others being zero. This minimum value is most often set equal to zero.}\BibitemShut {Stop}%
\bibitem [{\citenamefont {Nielsen}\ and\ \citenamefont {Kempe}(2001)}]{Nielsen2001}%
  \BibitemOpen
  \bibfield  {author} {\bibinfo {author} {\bibfnamefont {M.~A.}\ \bibnamefont {Nielsen}}\ and\ \bibinfo {author} {\bibfnamefont {J.}~\bibnamefont {Kempe}},\ }\bibfield  {title} {\bibinfo {title} {Separable states are more disordered globally than locally},\ }\href {https://doi.org/10.1103/PhysRevLett.86.5184} {\bibfield  {journal} {\bibinfo  {journal} {Phys. Rev. Lett.}\ }\textbf {\bibinfo {volume} {86}},\ \bibinfo {pages} {5184} (\bibinfo {year} {2001})}\BibitemShut {NoStop}%
\bibitem [{\citenamefont {Schneeloch}\ and\ \citenamefont {Howell}(2016)}]{schneeloch2016introduction}%
  \BibitemOpen
  \bibfield  {author} {\bibinfo {author} {\bibfnamefont {J.}~\bibnamefont {Schneeloch}}\ and\ \bibinfo {author} {\bibfnamefont {J.~C.}\ \bibnamefont {Howell}},\ }\bibfield  {title} {\bibinfo {title} {Introduction to the transverse spatial correlations in spontaneous parametric down-conversion through the biphoton birth zone},\ }\href@noop {} {\bibfield  {journal} {\bibinfo  {journal} {Journal of Optics}\ }\textbf {\bibinfo {volume} {18}},\ \bibinfo {pages} {053501} (\bibinfo {year} {2016})}\BibitemShut {NoStop}%
\bibitem [{\citenamefont {Kato}\ and\ \citenamefont {Takaoka}(2002)}]{kato2002sellmeier}%
  \BibitemOpen
  \bibfield  {author} {\bibinfo {author} {\bibfnamefont {K.}~\bibnamefont {Kato}}\ and\ \bibinfo {author} {\bibfnamefont {E.}~\bibnamefont {Takaoka}},\ }\bibfield  {title} {\bibinfo {title} {Sellmeier and thermo-optic dispersion formulas for ktp},\ }\href@noop {} {\bibfield  {journal} {\bibinfo  {journal} {Applied optics}\ }\textbf {\bibinfo {volume} {41}},\ \bibinfo {pages} {5040} (\bibinfo {year} {2002})}\BibitemShut {NoStop}%
\bibitem [{\citenamefont {Ramsay}\ \emph {et~al.}(1940)\citenamefont {Ramsay}, \citenamefont {Koppius},\ and\ \citenamefont {Cleveland}}]{Ramsay:40}%
  \BibitemOpen
  \bibfield  {author} {\bibinfo {author} {\bibfnamefont {B.~P.}\ \bibnamefont {Ramsay}}, \bibinfo {author} {\bibfnamefont {O.~T.}\ \bibnamefont {Koppius}},\ and\ \bibinfo {author} {\bibfnamefont {E.~L.}\ \bibnamefont {Cleveland}},\ }\bibfield  {title} {\bibinfo {title} {The prism and the theory of optical resolution},\ }\href {https://doi.org/10.1364/JOSA.30.000439} {\bibfield  {journal} {\bibinfo  {journal} {J. Opt. Soc. Am.}\ }\textbf {\bibinfo {volume} {30}},\ \bibinfo {pages} {439} (\bibinfo {year} {1940})}\BibitemShut {NoStop}%
\bibitem [{Note5()}]{Note5}%
  \BibitemOpen
  \bibinfo {note} {For an equilateral prism of side length $b$, the Rayleigh resolving power of a prism (surrounded by vacuum) may be given by the formula $\Delta \lambda /\lambda \approx -b n(n-n_{g})/\lambda $.}\BibitemShut {Stop}%
\bibitem [{\citenamefont {Starling}\ \emph {et~al.}(2012)\citenamefont {Starling}, \citenamefont {Bloch}, \citenamefont {Vudyasetu}, \citenamefont {Choi}, \citenamefont {Little},\ and\ \citenamefont {Howell}}]{AtomicPrismPRA2012}%
  \BibitemOpen
  \bibfield  {author} {\bibinfo {author} {\bibfnamefont {D.~J.}\ \bibnamefont {Starling}}, \bibinfo {author} {\bibfnamefont {S.~M.}\ \bibnamefont {Bloch}}, \bibinfo {author} {\bibfnamefont {P.~K.}\ \bibnamefont {Vudyasetu}}, \bibinfo {author} {\bibfnamefont {J.~S.}\ \bibnamefont {Choi}}, \bibinfo {author} {\bibfnamefont {B.}~\bibnamefont {Little}},\ and\ \bibinfo {author} {\bibfnamefont {J.~C.}\ \bibnamefont {Howell}},\ }\bibfield  {title} {\bibinfo {title} {Double lorentzian atomic prism},\ }\href {https://doi.org/10.1103/PhysRevA.86.023826} {\bibfield  {journal} {\bibinfo  {journal} {Phys. Rev. A}\ }\textbf {\bibinfo {volume} {86}},\ \bibinfo {pages} {023826} (\bibinfo {year} {2012})}\BibitemShut {NoStop}%
\bibitem [{\citenamefont {Shirasaki}(1999)}]{VIPAPatent}%
  \BibitemOpen
  \bibfield  {author} {\bibinfo {author} {\bibfnamefont {M.}~\bibnamefont {Shirasaki}},\ }\href@noop {} {\bibinfo {title} {Virtually imaged phased array as a wavelength demultiplexer}} (\bibinfo {year} {U.S. Patent 5 999 320, Dec. 7, 1999})\BibitemShut {NoStop}%
\bibitem [{\citenamefont {Xiao}\ \emph {et~al.}(2004)\citenamefont {Xiao}, \citenamefont {Weiner},\ and\ \citenamefont {Lin}}]{xiao2004dispersion}%
  \BibitemOpen
  \bibfield  {author} {\bibinfo {author} {\bibfnamefont {S.}~\bibnamefont {Xiao}}, \bibinfo {author} {\bibfnamefont {A.~M.}\ \bibnamefont {Weiner}},\ and\ \bibinfo {author} {\bibfnamefont {C.}~\bibnamefont {Lin}},\ }\bibfield  {title} {\bibinfo {title} {A dispersion law for virtually imaged phased-array spectral dispersers based on paraxial wave theory},\ }\href@noop {} {\bibfield  {journal} {\bibinfo  {journal} {IEEE journal of quantum electronics}\ }\textbf {\bibinfo {volume} {40}},\ \bibinfo {pages} {420} (\bibinfo {year} {2004})}\BibitemShut {NoStop}%
\bibitem [{\citenamefont {Xiao}\ and\ \citenamefont {Weiner}(2004)}]{xiao20042}%
  \BibitemOpen
  \bibfield  {author} {\bibinfo {author} {\bibfnamefont {S.}~\bibnamefont {Xiao}}\ and\ \bibinfo {author} {\bibfnamefont {A.~M.}\ \bibnamefont {Weiner}},\ }\bibfield  {title} {\bibinfo {title} {2-d wavelength demultiplexer with potential for $\geq$ 1000 channels in the c-band},\ }\href@noop {} {\bibfield  {journal} {\bibinfo  {journal} {Optics Express}\ }\textbf {\bibinfo {volume} {12}},\ \bibinfo {pages} {2895} (\bibinfo {year} {2004})}\BibitemShut {NoStop}%
\bibitem [{\citenamefont {Zhu}\ \emph {et~al.}(2020)\citenamefont {Zhu}, \citenamefont {Lin}, \citenamefont {Hao}, \citenamefont {Wang},\ and\ \citenamefont {He}}]{Zhu_2020}%
  \BibitemOpen
  \bibfield  {author} {\bibinfo {author} {\bibfnamefont {X.}~\bibnamefont {Zhu}}, \bibinfo {author} {\bibfnamefont {D.}~\bibnamefont {Lin}}, \bibinfo {author} {\bibfnamefont {Z.}~\bibnamefont {Hao}}, \bibinfo {author} {\bibfnamefont {L.}~\bibnamefont {Wang}},\ and\ \bibinfo {author} {\bibfnamefont {J.}~\bibnamefont {He}},\ }\bibfield  {title} {\bibinfo {title} {A vipa spectrograph with ultra-high resolution and wavelength calibration for astronomical applications},\ }\href {https://doi.org/10.3847/1538-3881/aba836} {\bibfield  {journal} {\bibinfo  {journal} {The Astronomical Journal}\ }\textbf {\bibinfo {volume} {160}},\ \bibinfo {pages} {135} (\bibinfo {year} {2020})}\BibitemShut {NoStop}%
\end{thebibliography}%

\newpage
\appendix
\begin{widetext}

\section{Proof that coarse-graining cannot increase relative entropy}
To illustrate why continuous majorization is useful, we present an expanded version of the proof in the supplementary material of \cite{schneeloch2019quantifying} that majorization cannot increase the relative entropy between two probability distributions.

Consider two discrete probability distributions $P_{1}$ and $Q_{1}$ each with $n$ possible outcomes. The entropy of $Q_{1}$ relative to $P_{1}$, denoted $\mathscr{D}(P_{1}||Q_{1})$ is given by:
\begin{equation}
\mathscr{D}(P_{1}||Q_{1})\equiv \sum_{i}P_{1i}\log\left(\frac{P_{1i}}{Q_{1i}}\right)
\end{equation}
Here and throughout the paper, we assume logarithms to be base 2, since we measure entropy in bits.

Let us assume two new probability distributions $P_{2}$ and $Q_{2}$ obtained from $P_{1}$ and $Q_{1}$ through the same permutation operation. Since the relative entropy is invariant under simultaneous permutations of outcomes, we have:
\begin{equation}
\mathscr{D}(P_{1}||Q_{1}) = \mathscr{D}(P_{2}||Q_{2})
\end{equation}
The relative entropy is a jointly convex function \cite{Cover2006}. Because of this, the majorization operation that comes from mixing permutations of a probability distribution cannot increase relative entropy:
\begin{align}
\mathscr{D}(\lambda P_{1} &+ (1-\lambda)P_{2}||\lambda Q_{1} + (1-\lambda)Q_{2})\nn\\
&\leq\lambda \mathscr{D}(P_{1}||Q_{1}) + (1-\lambda) \mathscr{D}(P_{2}||Q_{2})\nn\\
&\leq \min\{\mathscr{D}(P_{1}||Q_{1}),\mathscr{D}(P_{2}||Q_{2})\}
\end{align}

Note that multiple forms of entropy can be expressed in terms of relative entropy:
\begin{align}
\mathscr{D}(P(X_{A})||P(U_{A}))&=\log(n_{A})-H(X_{A})\\
\mathscr{D}(P(X_{A},X_{B})||P(U_{A})P(U_{B}))&=\log(n_{A}n_{B}) -H(X_{A},X_{B})\\
\mathscr{D}(P(X_{A},X_{B})||P(U_{A})P(X_{B}))&=\log(n_{A}) -H(X_{A}|X_{B})\\
\mathscr{D}(P(X_{A},X_{B})||P(X_{A})P(X_{B}))&=H(X_{A}:X_{B})
\end{align}
Here, $P(U_{A})$ and $P(U_{B})$ are the uniform probability distributions over the spaces defining the variables $X_{A}$ and $X_{B}$, respectively. To prove that majorization cannot increase relative entropy, we use repeated applications of its joint convexity. 

Let us define the mixed distributions $\bar{P}=\lambda P_{1}+(1-\lambda)P_{2}$, and $\bar{Q}$ similarly for $\lambda\in[0,1]$. By use of the joint convexity of relative entropy, we may show the following:
\begin{align}
\mathscr{D}(P(X,Y)||P(X)P(Y)) \geq \mathscr{D}(\bar{P}(X,Y)||P(X)P(Y))\nn\\
\geq \mathscr{D}(\bar{P}(X,Y)||\bar{P}(X)P(Y))\nn\\
\geq \mathscr{D}(\bar{P}(X,Y)||\bar{P}(X)\bar{P}(Y))
\end{align}
which proves that majorization cannot increase relative entropy. Consequently, we have that majorization cannot decrease joint entropy $H(X_{A},X_{B})$, marginal entropy $H(X_{A})$, or conditional entropy $H(X_{A}|X_{B})$, and cannot increase mutual information $H(X_{A}:X_{B})$ or even conditional mutual information $H(X_{A}:X_{B}|X_{C})$, which can also be expressed in terms of relative entropy.

\end{widetext}
\end{document}